\documentclass[nonatbib]{article}

\usepackage[final]{tackling_climate_workshop_style}

\usepackage{graphicx}
\usepackage[utf8]{inputenc} 
\usepackage[T1]{fontenc}    
\usepackage[colorlinks=true]{hyperref}       
\usepackage{url}            

\usepackage{booktabs}       
\usepackage{amsfonts}       
\usepackage{nicefrac}       
\usepackage{microtype}      

\title{Stress-testing the coupled behavior of hybrid physics-machine learning climate simulations on an unseen, warmer climate}
\author{Jerry Lin \\
        University of California, Irvine\\
  \texttt{jerryL9@uci.edu}
  \And
  Mohamed Aziz Bhouri \\
  Columbia University \\
  \texttt{mb4957@columbia.edu}
  \And
  Tom Beucler \\
  University of Lausanne \\
  \texttt{tom.beucler@unil.ch}
  \And
  Sungduk Yu \\
  University of California, Irvine \\
  \texttt{sungduk@uci.edu}
  \And
  Michael Pritchard \\
  University of California, Irvine and NVIDIA\\
  \texttt{mspritch@uci.edu}
    }
\begin{document}

\maketitle

\begin{abstract}

Accurate and computationally-viable representations of clouds and turbulence are a long-standing challenge for climate model development. Traditional parameterizations that crudely but efficiently approximate these processes are a leading source of uncertainty in long-term projected warming and precipitation patterns. Machine Learning (ML)-based parameterizations have long been hailed as a promising alternative with the potential to yield higher accuracy at a fraction of the cost of more explicit simulations. However, these ML variants are often unpredictably unstable and inaccurate in \textit{coupled} testing (i.e. in a downstream hybrid simulation task where they are dynamically interacting with the large-scale climate model). These issues are exacerbated in out-of-distribution climates. Certain design decisions such as ``climate-invariant" feature transformation for moisture inputs, input vector expansion, and temporal history incorporation have been shown to improve coupled performance, but they may be insufficient for coupled out-of-distribution generalization. If feature selection and transformations can inoculate hybrid physics-ML climate models from non-physical, out-of-distribution extrapolation in a changing climate, there is far greater potential in extrapolating from observational data. Otherwise, training on multiple simulated climates becomes an inevitable necessity. While our results show generalization benefits from these design decisions, the obtained improvment does not sufficiently preclude the necessity of using multi-climate simulated training data.
  
\end{abstract}

\section{Introduction and Motivation}

Anthropogenic climate change is increasing the frequency and severity of climate extremes and natural disasters, requiring informed adaptation and mitigation measures from policymakers \cite{IPCC2021, Donat2017-qq, Vargas_Zeppetello2022-pp}. While domain scientists continue to achieve notable progress in improving our climate physics understanding, significant uncertainty in projected warming and precipitation patterns remains. Much of this uncertainty stems from the intractable computational expense of explicitly resolving subgrid processes like convection and radiation, making cheaper \textit{conventional} parameterizations that approximate their effects necessary \cite{IPCC2021, Tian2020-jp}. Even if hardware advancements continue at the pace of Moore's Law, it would take decades to be able to run global climate simulations that resolve the turbulent eddies responsible for low cloud formation, a major source of uncertainty in projected warming \cite{Schneider2017-od}. 

Neural network parameterizations could be trained on more explicit simulations to emulate unresolved subgrid processes, enabling a higher fidelity representation using current-generation hardware \cite{Brenowitz2018, Gentine2018-ux, OGorman2018, Mooers2021-sh, Wang2022-po, Yu2023-sv}. However, the task of parameterizing subgrid physics (in our case convection and radiation) becomes stubbornly difficult when these neural network emulators are coupled to the large-scale climate model and integrated in time. Because coupled behavior is highly variable, large-scale coupled tests are necessary for drawing conclusions on surrogate model design decisions \cite{Lin2023-he}. Based on proven coupled in-distribution benefits \cite{Lin2023-he} and using large-scale coupled tests on an out-of-distribution climate, we rigorously test generalization improvement of the following three design decisions:

\begin{enumerate}
\item Using a relative humidity ``climate-invariant" feature transformation for the moisture input \cite{Beucler2021-ai}.
\item Expanding the input vector to address potential omitted variable bias \cite{Lin2023-he}
\item Incorporating memory effects (i.e. temporal history) in the input \cite{Han2020-xy, Lin2023-he}
\end{enumerate}

Our results show that these design decisions improve generalization on an out-of-distribution climate relative to our baseline neural network configuration, but they are not sufficient to supplant multi-climate training, something that is argued necessary in previous works \cite{Clark2022-uq, Bhouri2023-MF}.
 
\section{Methods}

\subsection{Reference Climate Simulation}

Our neural networks are trained on and validated against the Super-Parameterized Community Atmosphere Model v3 (SPCAM 3), which has served as a test-bed of for prototyping neural network emulators of subgrid convection in previous works \cite{Gentine2018-ux, Rasp2018-fk, Ott2020-qe, Beucler2021-ai, Behrens2022-qq}. In super-parameterization, a high-fidelity, 2D model of convection called a Cloud-Resolving Model (CRM) with 32 columns at a 4-km horizontal resolution is embedded inside each grid-cell of the host climate model \cite{Khairoutdinov2005-xq, Pritchard2014-cv, Jones2019-ke}. For simplicity, we use a fixed season, prescribed sea surface temperatures, and a zonally-symmetric aquaplanet. The timestep is fixed to 30 minutes and 30 vertical levels are considered within each grid cell. Using SPCAM 3, we create two reference simulations, with one having prescribed sea surface temperatures that are 4K warmer. All hybrid-ML climate models are trained using the colder climate and coupled in both settings. 

\subsection{Neural Network Configurations}

To assess the generalization benefits of our design decisions, we train and evaluate 330 neural networks for each design decision and each of the following configurations. The Specific Humidity (SH) Configuration is a baseline similar to previous work \cite{Rasp2018-fk}. The Relative Humidity (RH) Configuration uses a ``climate-invariant" relative humidity feature transformation for the moisture input variable. The Expanded Variables (EV) Configuration concatenates meridional wind, ozone mixing ratio, and cosine of zenith angle to the input variables of the RH configuration. Finally, the Previous Tendencies (PT) Configuration concatenates heating and moistening tendencies from one previous timestep to the input variables of the RH configuration. In order to leverage scalable off-the-shelf tools for coupling (in our case the Fortran Keras Bridges (FKB)), all neural networks are dense, feedforward neural networks. The input and output variables and hyperparameter search space used for sampling can be found in the Appendix.

\section{Results}

\subsection{Uncoupled Results}

\begin{figure}
  \centering
  \includegraphics[width=\textwidth]{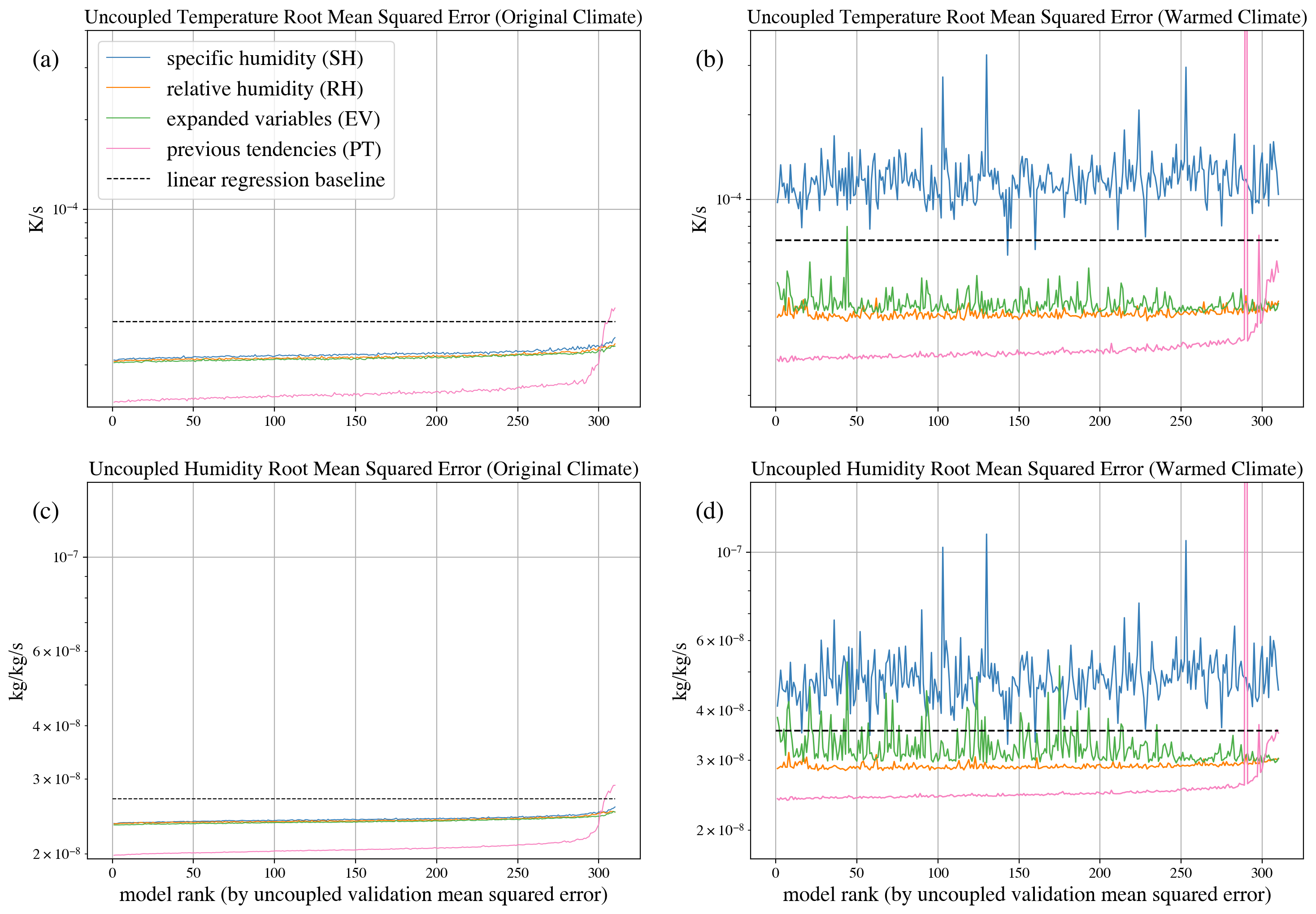}
  \caption{Uncoupled test error on in-distribution and out-of-distribution (warmer) climate for each configuration with models ranked by validation error.}
  \label{fig:uncoupled_error}
\end{figure}

The uncoupled error for all four configurations in both climates is given in Figure \ref{fig:uncoupled_error}, showing higher values across the board in the warmer climate. As expected, the baseline SH configuration no longer clears the linear regression baseline when tested on the warmer climate, in line with smaller-scale uncoupled results from Beucler et al. (2021) \cite{Beucler2021-ai}. Variance in uncoupled error for the SH (EV) configuration jumps by 2 (3) orders of magnitude in the warmer climate testing, and error from the EV configuration is higher compared to the RH configuration (a reversal from uncoupled results in the original climate), suggesting potential overfitting on some of the additional variables.

\subsection{Coupled Results}

\subsubsection{Coupled Generalization on Unseen Climate}

\begin{figure}
  \centering
  \includegraphics[width=\textwidth]{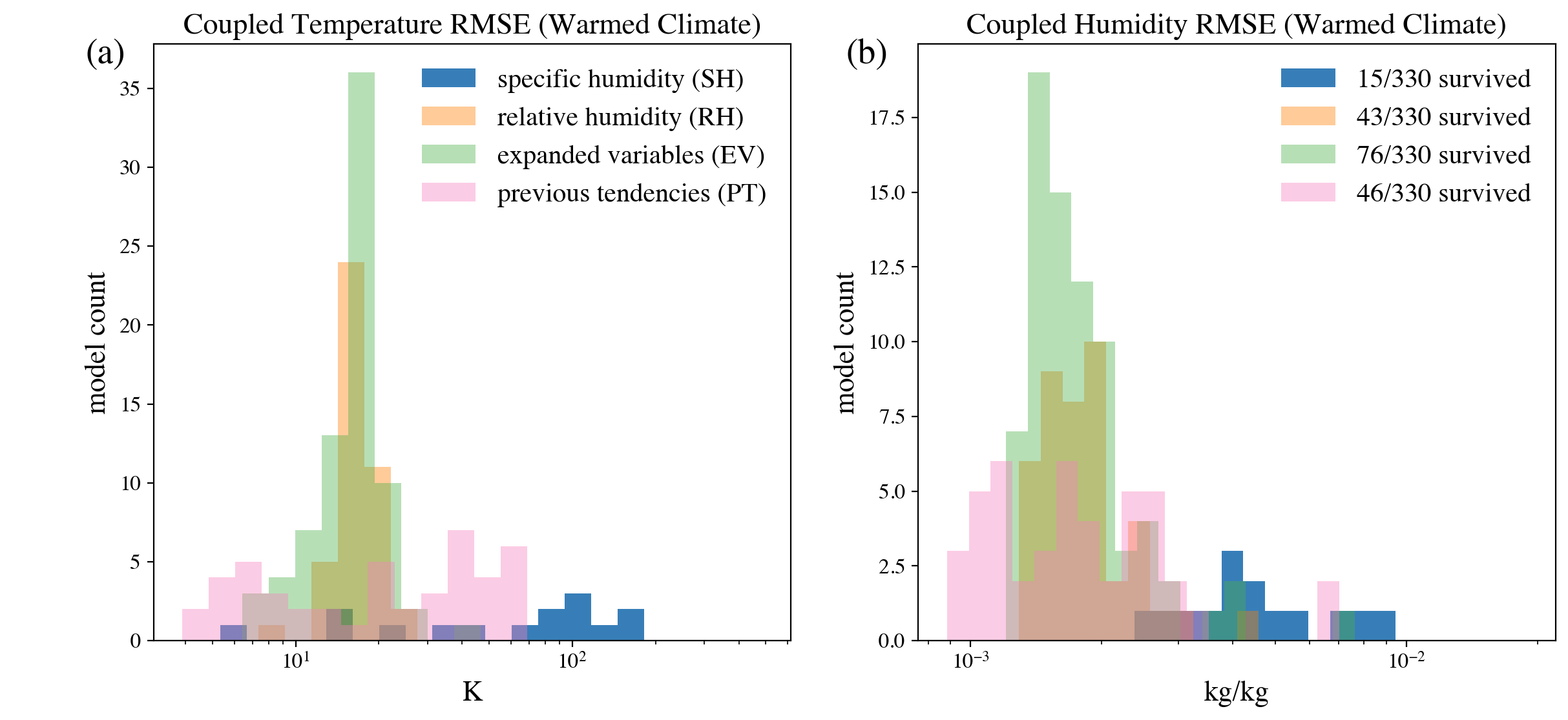}
  \caption{Histograms for coupled root mean squared error for temperature and humidity in the warmer (unseen) climate. Models corresponding to coupled simulations that prematurely terminated are excluded.}
  \label{fig:coupled_p4k_error}
\end{figure}

Figure \ref{fig:coupled_p4k_error} shows the coupled results for the warmer climate testing, highlighting potential success in coupled generalization to an unseen climate. However, only a fraction of the coupled simulations run for the entire year without prematurely terminating (4.5\%, 13\%, 23\%, and 14\% of simulations for the SH, RH, EV and PT configurations, respectively). While the coupled simulations with the lowest temperature and humidity RMSEs belong to the PT configuration, the best PT models for one climate are not necessarily the best for the other.

\begin{figure}
  \centering
  \includegraphics[width=\textwidth]{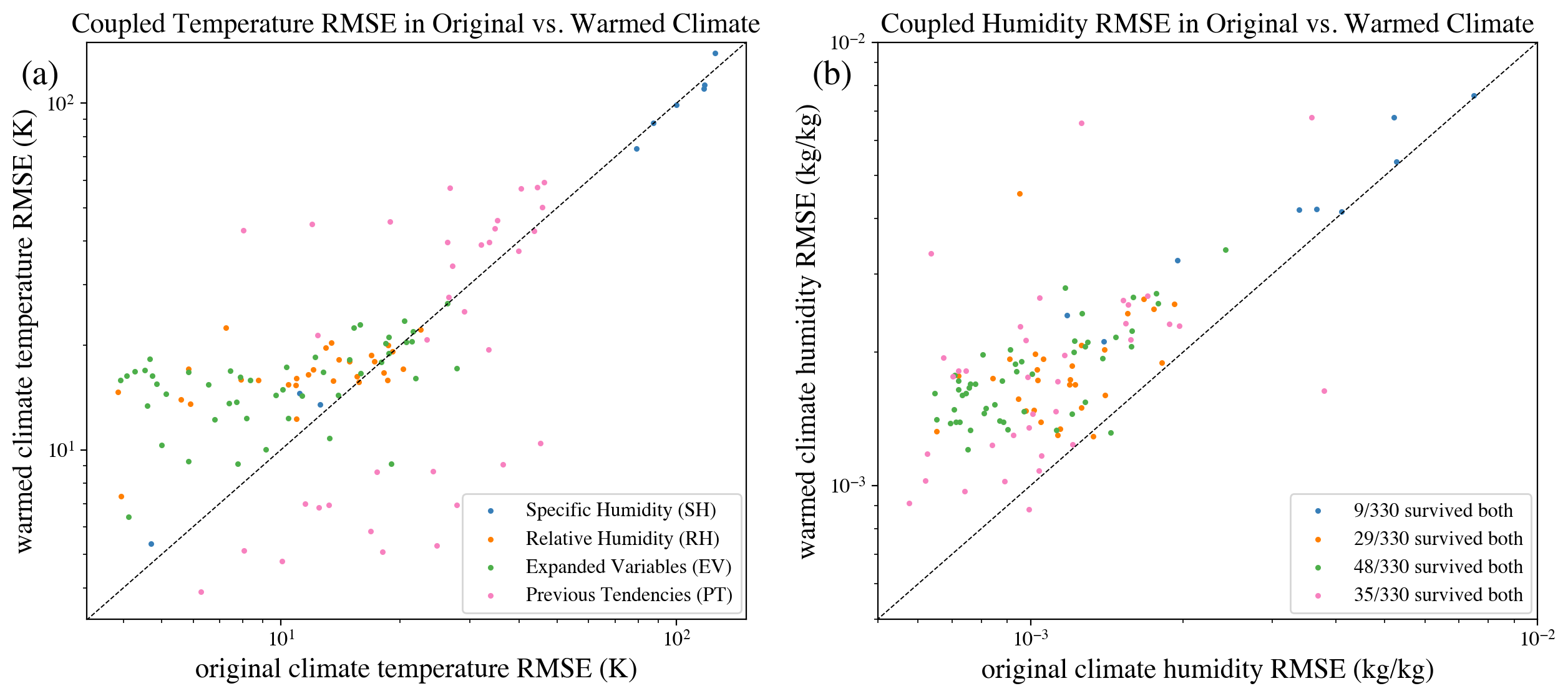}
  \caption{Scatterplots of coupled error for hybrid physics-ML simulations that did not prematurely terminate in either climate. Dashed line is a 1-to-1 line that intersects the origin.}
  \label{fig:originalvwarmed}
\end{figure}

\subsubsection{Coupled Generalization in Both Climates}

Figure \ref{fig:originalvwarmed} shows a fairly clear relationship between coupled error in the warmer climate and the original one, with $R^2$ values of .82 and .61 for temperature and humidity, respectively. However, excluding the SH configuration, whose models under-performed linear regression in the warmer climate, drops these $R^2$ values to .4 and .25, respectively.

\section{Discussion}

As seen in Figure \ref{fig:originalvwarmed}, there appears to be a weak relationship for coupled temperature error between in-distribution and out-of-distribution climates. Such a relationship is not only much weaker for moisture results, but the coupled moisture errors in the warmer climate are almost uniformly higher. This behavior indicates that generalizing on moistening in a coupled setting on an out-of-distribution climate deserves focus for future model development. It is possible that addressing this limitation in future model development will also tighten the relationship for temperature error between in-distribution and out-of-distribution climates.

\section{Conclusion}

Coupled out-of-distribution generalization might still be possible without multi-climate training when using more sophisticated network architectures, physics-informed neural networks (PINNs), pruned feature selection, and additional ``climate-invariant" feature transformations (e.g. for temperature and latent heat flux) \cite{Beucler2019-zz, Iglesias-Suarez2023-qm, Beucler2021-ai, Lin2023-he}. It is also worth noting that enforcing conservation laws for our task is not possible without additional inputs and outputs. Nevertheless, our results point to the necessity of stress-testing out-of-distribution in a coupled setting. The combined task of generalizing out-of-distribution and remaining stable and accurate when coupled is a demonstrably more difficult challenge that requires further collaboration between domain scientists and machine learning experts.

\begin{ack}

We would like to thank Justus Will and Ritwik Gupta for helpful feedback on this manuscript. High-performance computing was facilitated by Bridges\-2 at the Pittsburgh Supercomputing Center through allocation ATM190002 from the Advanced Cyberinfrastructure Coordination Ecosystem: Services \& Support (ACCESS) program, which is supported by NSF grants \#2138259, \#2138286, \#2138307, \#2137603, and \#2138296. J.L., S.Y., and M.S.P acknowledge support from the DOE (DE-SC0023368) and NSF (AGS-1912134). M.A.B acknowledges National Science Foundation funding from an AGS-PRF Fellowship Award (AGS2218197). Finally, we would like to thank Xiaojuan Liu for donating unused CPU hours on Bridges\-2. 



\end{ack}

\bibliographystyle{ieeetr}
\bibliography{tackling_climate_workshop}

\subsection{Input and output variables}

Table \ref{tab:input} depicts the input variables for the neural network configurations. All neural network configurations other than the baseline specific humidity (SH) configuration use relative humidity (\%) for moisture. Variables marked with a single asterisk * are exclusive to the expanded variables (EV) configuration, and variables marked with a double asterisk ** are exclusive to the previous tendencies configuration. Input variables are normalized by subtracting the mean and dividing by the standard deviation.

All neural network configurations share output variables shown in \ref{tab:output}. Output variables are multiplied by 1004 and 2.5e6 for heating and moistening tendencies, respectively, to put them in similar orders of magnitude.

\begin{table}[ht]
\caption{Input variables}
\label{tab:input}
\centering
\begin{tabular}{l l l l}
\toprule
  Input variable  & Unit  & \\
\midrule
   Temperature  & $K$ & 30 \\
   Humidity  & $kg/kg$ or \% & 30\\
   Surface pressure  & $Pa$  & 1 \\
   Incoming solar radiation  & $W/m^2$  & 1 \\
   Sensible heat flux  & $W/m^2$ & 1 \\
   Latent heat flux  & $W/m^2$ & 1  \\
   Meridional wind*  & $m/s$ & 30 \\
   Ozone mixing ratio*  & $m^3/m^3$ & 30 \\
   Cosine of zenith angle*  &  & 1 \\
   $(t-1)$ Heating tendency** & $K/s$  & 30 \\
   $(t-1)$ Moistening tendency**  & $kg/kg/s$  & 30\\
\bottomrule
\\
\end{tabular}
\end{table}

\begin{table}[ht]
\caption{Output variables}
\label{tab:output}
\centering
\begin{tabular}{l l l l}
\toprule
  Input variable  & Unit  & Vertical levels\\
\midrule
   Heating tendency  & $K/s$ & 30 \\
   Moistening tendency  & $kg/kg/s$ & 30 \\
\bottomrule
\\
\end{tabular}
\end{table}

\subsection{Hyperparameter search space}

All neural networks uniformly randomly subsample the search space depicted in Table \ref{tab:hp_search}. For the learning rate range, the entire interval is log transformed such that different orders of magnitude are sampled at similar rates. 

\begin{table}[ht]
\caption{Hyperparameter search space}
\label{tab:hp_search}
\centering
\begin{tabular}{l l}
\toprule
  Hyperparameter  & Range  \\
\midrule
   Hidden layers  & [[4, 11]]  \\
   Nodes per layer  & [[128, 512]]  \\
   Batch normalization  & \{On, Off\}   \\
   Dropout  & [0.0, 0.25]   \\
   Optimizer  & \{RMSprop, Adam, RAdam, QHAdam\}  \\
   Leaky ReLu slope  & [0.0, 0.4]   \\
   Learning rate  & [1e-5, 1e-2]  \\
\bottomrule
\\
\end{tabular}

\end{table}







\end{document}